# Ferroelectric polarization switching with a remarkably high activation-energy in orthorhombic GaFeO$_3$ thin films

Running title : Polarization switching in o-GFO thin film at RT

Seungwoo Song,[1] Hyun Myung Jang,[1*] Nam-Suk Lee,[2] Jong Y. Son,[3] Rajeev Gupta,[4] Ashish Garg,[4] Jirawit Ratanapreechachai,[5] and James F. Scott[5,6+]

[1]Division of Advanced Materials Science, and Department of Materials Science and Engineering, Pohang University of Science and Technology (POSTECH), Pohang 790-784, Republic of Korea.  [2]National Institute for Nanomaterials Technology, Pohang University of Science and Technology (POSTECH), Pohang 790-784, Republic of Korea.

[3]Department of Applied Physics, College of Applied Science, Kyung Hee University, Giheung-Gu, Yongin-City 446-701, Republic of Korea.

[4]Department of Materials Science and Engineering, Indian Institute of Technology (IIT), Kanpur, Kanpur 208016, India.

[5]Cavendish Laboratory, Department of Physics, University of Cambridge, Cambridge CB3 0HE, UK.  [6]School of Chemistry and School of Physics, University of St. Andrews, St. Andrews KY16 9ST, UK.

Correspondence and requests for materials should be addressed to either H. M. J. (email: hmjang@postech.ac.kr ) or J. F. S. (email: jfs32@cam.ac.uk )

Orthorhombic GaFeO$_3$ (o-GFO) with the polar Pna2$_1$ space group is a prominent ferrite by virtue of its piezoelectricity and ferrimagnetism, coupled with magneto-electric effects. Herein, we unequivocally demonstrate a large ferroelectric remanent polarization in undoped o-GFO thin films by adopting either a hexagonal strontium titanate (STO) or a cubic yttrium-stabilized zirconia (YSZ) substrate. The polarization-electric-field hysteresis curves of the polar c-axis-grown o-GFO film on a SrTiO$_3$/STO substrate show the net switching polarization of ~35 $\mu$C/cm$^2$ with an unusually high coercive field of $\pm$1400 kV/cm at room temperature. The PUND measurement also demonstrates the switching polarization of ~26 $\mu$C/cm$^2$. The activation energy for the polarization switching, as obtained by density-functional theory calculations, is remarkably high, 1.05 eV per formula unit. This high value accounts for the observed stability of the polar Pna2$_1$ phase over a wide range of temperature up to 1368 K.

**INTRODUCTION**

Multiferroics are an interesting group of materials that simultaneously exhibit ferroelectricity and magnetic ordering with coupled electric, magnetic, and structural orders. Multiferroic materials with a pronounced degree of magnetoelectric (ME) coupling at room temperature are of great scientific and technological importance for their use in various types of electronic devices that include sensors, actuators, and electric-field controllable magnetic memories.[1-3] Among all the known multiferroics, $BiFeO_3$ is most extensively studied owing to its large room-temperature spontaneous polarization with improved magnetic properties in epitaxially strained thin-film forms.[4-7]

Polar orthorhombic $GaFeO_3$ (*o*-GFO) is another prominent multiferroic oxide by virtue of its room-temperature piezoelectricity (possibly ferroelectricity as well), near room-temperature ferrimagnetism, and pronounced low-temperature ME effects. Since a linear ME effect was first reported in 1960s by Rado,[8] magnetization-induced second harmonic generation,[9] optical ME effect,[10] and other interesting studies[11-22] keep renewing our research attention to this system. GFO crystallizes into the polar orthorhombic $Pna2_1$ (equivalently, $Pc2_1n$) space group with a ferrimagnetically ordered spin structure. The four $Fe^{3+}$ ions in a unit cell are antiferromagnetically coupled along *a*-axis in the $Pna2_1$ setting. However, the intermixed $Fe^{3+}$ ions occupying at the Ga sites (in other words, different Fe occupations at the Fe1 and Fe2 sites) can lead to a ferromagnetic order with $Fe^{3+}$ ions at the Fe sites, which gives rise to ferrimagnetic ordering at a temperature around 230 K.[11,12]

Contrary to ferrimagnetic ordering and low-temperature ME effects,[11] much less is known on the ferroelectricity of *o*-GFO. In principle, *o*-GFO should exhibit ferroelectricity up to high temperatures as the polar $Pna2_1$ phase (with the corresponding point group of 2mm or *C2v* in Schönflies notation) remains stable, at least, up to 1368 K.[22] According to Stoeffler,[20,23] the computed *ab initio* polarization value of the undoped *o*-GFO is as high as 25 $\mu$C/cm$^2$. In spite of these predictions, however, there has been no experimental demonstration of the room-temperature ferroelectric polarization switching with its polarization value compatible with the *ab initio* predictions. Several groups have studied the *P-E* (polarization-electric-field) response of GFO ferroelectrics, mostly using thin-film forms.[13-16] However, all these studied revealed that the remanent polarization ($P_r$) of GFO is less than 0.5 $\mu$C/cm$^2$ with an unrealistically small value of the coercive field ($E_c \leq 5$ kV/cm) as well. More recently, Oh *et al.*[17] reported the epitaxial growth of *o*-GFO film on a SrRuO$_3$(111)/SrTiO$_3$(111) substrate with the remarkably enhanced $E_c$ of ~150 kV/cm. However, their $P_r$ value still remains at ~0.05 $\mu$C/cm$^2$. On the other hand, Mukherjee *et al.*[18] recently reported room-temperature ferroelectricity of the *o*-GFO thin film grown on an ITO(001)//YSZ(001) substrate on the basis of their observation of an 180$^o$ phase-shift of piezoresponse. However, the 180$^o$ phase-shift (or switching of the piezoelectric phase) does not necessarily indicate a ferroelectric polarization switching across the barrier of a double-well potential.

Up to now, the only unequivocal experimental demonstration of a reversible polarization switching in *o*-GFO thin films is made by Thomasson *et al.*[19] According to their study, the 2 % Mg-doped GFO film exhibits a well saturated *P-E* switching curve with a negligible tendency of the non-switching polarization.[19] However, the measured $P_r$ value is as small as 0.2 $\mu$C/cm$^2$,[19] which is less than 1 % of the predicted $P_r$ value (25 $\mu$C/cm$^2$) of the undoped GFO. Moreover, this system is not a stoichiometric *o*-GFO (i.e., not an orthoferrite with Ga:Fe = 0.6:1.4), in addition to 2% Mg-doping.[19] Thus, all the reported $P_r$ values of *o*-GFO are in the range of 0.05 and 0.5 $\mu$C/cm$^2$, which is unacceptably too small to be compatible with the *ab* *initio* polarization of 25 $\mu C/cm^2$.[20,23]

In this article, we have clarified a puzzling discrepancy between the observed $P_r$ value ($\leq$ 0.5 $\mu C/cm^2$) and the *ab initio* prediction (25 $\mu C/cm^2$) and unequivocally demonstrated the ferroelectric polarization switching with the net switching polarization value of ~30 $\mu C/cm^2$ by using the *o*-GFO thin films preferentially grown along the polar *c*-axis in *Pna*$2_1$ setting (*b*-axis in *Pc*$2_1$*n* setting). In the present study, we adopt either a hexagonal or a cubic substrate to demonstrate the room-temperature polarization switching**:** (i) a SrRuO$_3$ (SRO) (111) buffered hexagonal strontium titanate (STO) (111) substrate and (ii) an ITO (001) buffered cubic yttrium-stabilized zirconia (YSZ) (001) substrate.

## MATERIALS AND METHODS

**Experimental methods.** Orthorhombic GaFeO$_3$ (*o*-GFO) film and SRO bottom electrode were grown on a Ti$^{4+}$-single-terminated STO (111) substrate by pulsed laser deposition with KrF excimer laser ($\lambda$ = 248nm) operated at 3 Hz and 10 Hz, respectively. GFO films were deposited at 800 ºC in an oxygen ambient atmosphere (200 mTorr) with a fluence of 1 J/cm$^2$ focusing on a stoichiometrically sintered *o*-GFO target while 30-nm-thin SRO bottom-electrode layers were grown at 680 ºC in a 100 mTorr oxygen atmosphere with a fluence of 2 J/cm$^2$. After the deposition, the SRO layer was cooled down to room temperature under the same oxygen pressure used in the *o*-GFO film deposition. We observed the thickness fringes of SRO(222) around the two-theta (2$\theta$) value of STO (222), which strongly indicates an epitaxial growth of the SRO bottom electrode layer. The thickness of this SRO layer (30 nm) was also determined from the positions of the interference fringes (See Supplementary Information for details).

Analysis of the domain orientation and phase formation in the films was done by using a high-resolution x-ray diffractometer (D8 discover, Bruker) under Cu Kα radiation. Domain structures were investigated in detail by employing high-resolution transmission electron microscopy (JEM-2100F, JEOL with a probe Cs-corrector). Z-contrast high-angle annular dark-field STEM (HAADF-STEM) image and selected-area electron diffraction (SAED) experiments were carried out under 200-kV acceleration voltage.

For dielectric-ferroelectric measurements of the *c*-axis-grown *o*-GFO film, the Pt top electrode with the diameter of 100 μm was deposited using a dc sputtering system. Current and voltage (I-V) curves were recorded using a Keithley 2400 source meter. Polarization-electric-field hysteresis loops (P-E curves) and positive-up and negative-down (PUND) pulse sequences were measured using a precision LC ferroelectric tester (Radiant technologies). A commercial atomic force microscope (DC-EFM in XE-100, Park Systems) was used for vertical piezoelectric force microscopy (vPFM) study to map piezoelectric properties of thin films. Pt/Ir coated tip was used for probing the signals. The input modulation voltage $V_{ac}$ (with the amplitude in the range of 0.1~2 V and the *ac* frequency of 15 kHz) was applied between the conductive tip and the bottom electrode using a function generator. The resulting oscillations of the cantilever are read out with a lock in amplifier. Magnetic properties were measured by superconducting quantum interference device magnetometry (SQUID, MPMS, Quantum Design) from 10 K to 300 K under various applied magnetic fields.

**Computational methods.** We have performed DFT calculations of o-GFO on the basis of the generalized gradient approximation (GGA)[24] and GGA+U method[25] implemented with projector augmented-wave (PAW) pseudopotential[26] using the *Vienna ab initio Simulation Package* (VASP).[27] All the DFT calculations were performed by adopting (i) a 11×6×6

Monkhorst-Pack *k*-point mesh[28] centered at the Γ-point, (ii) a 650-eV plane-wave cutoff energy, and (iii) the tetrahedron method with Blöchl corrections for the Brillouin-zone integrations.[29] The structural optimizations were performed for the 40-atom-cell which corresponds to an orthorhombic unit cell consisting of 8 formula units. The ions were relaxed until the Hellmann-Feynmann forces on them were less than 0.01 eV/Å. The Hubbard $U_{eff}$ of 4 eV and intra-atomic exchange parameter (*J*) of 0.89 eV for the Fe 3*d*-orbital were chosen on the basis of the previous work.[20]

**RESULTS**

**In-plane Domain Orientation of the Polar *c*-axis-grown Film**

To fabricate the polar *c*-axis grown GFO film with a simpler domain configuration, we have carefully chosen a hexagonal STO(111) substrate as an alternative to a cubic YSZ(001) substrate. For implementing this, we adopt a SRO(111) bottom electrode by considering its compatibility with the STO substrate and fatigue and imprinting resistance.[30,31] This scheme of the *o*-GFO film growth on a hexagonal substrate enables us to achieve a substantial simplification in the domain configuration, from twelve orientations [on a cubic YSZ(001) substrate[18]] to six in-plane orientations. Figure 1a shows that the [001]-oriented GFO thin film is preferentially grown on a SRO(111) buffered STO(111) substrate. The calculated *c*-axis parameter using this *θ-2θ* XRD pattern is 9.3996 Å which essentially coincides with the bulk *c*-axis parameter.[13] This suggests that the present GFO film is fully relaxed along the growth direction. For the [001]-oriented film grown on a STO(111), the most probable domain orientation that minimizes the in-plane lattice mismatch is represented by three in-plane domain configurations ($D_1, D_2$ and $D_3$), as schematically depicted in Figure 1b.

To examine the validity of the proposed domain configuration, we have measured in-plane XRD phi($\varphi$)-scan and the result is presented in Figure 1c. These $\varphi$-scan spectra were obtained by keeping the Bragg angle at (013) for o-GFO film (orange line) and at (110) and (100) for STO (blue and green lines, respectively). In the case of GFO (013), six peaks appear (orange color) and each peak is separated from the two neighboring GFO (013) peaks by 60°. On the other hand, as shown in Figure 1c, these six (013) peaks are separated from STO (110) and (100) peaks by 30°. These two observations clearly indicate that the GFO [013] has six in-plane orientations with a successive tilting angle of 60° when projected on a STO (111) surface and each of these six in-plane orientations is rotated by 30° from the STO (111) projection of the STO [110] and [100] vectors. The domain configuration depicted in Figure 1d satisfies all these orientation relationships. The six projected normal vectors of the GFO (013) are shown using orange color in the lower corner of Figure 1d. Thus, the o-GFO film grown on the STO (111) surface is characterized by three distinct crystallographic variants, $D_1, D_2$ and $D_3$, and is represented by total six in-plane orientations, namely, $D_1, D_1', D_2, D_2'$ and $D_3, D_3'$, where $D_n$ and $D_n'$ are facing to each other with the same crystallographic variant. This domain configuration [Figure 1d] agrees well with our previous proposition deduced from minimizing the lattice mismatch [Figure 1b]. On the contrary, the *o*-GFO film grown on an ITO/YSZ(001) substrate is characterized by total twelve in-plane orientations with a successive in-plane tilting angle of 30° (See Supplementary Information for details).

However, there exists a considerable degree of the lattice mismatch (Δ) even though the (001)-oriented GFO film possesses six in-plane orientations to reduce the lattice mismatch between the film and substrate. According to our estimate, $\Delta_a$= 5.8654% ($a_{GFO} = 5.0806$Å, $\sqrt{2}a_{STO}\cos(30^o)$) and $\Delta_b$ = 5.3388% ($b_{GFO}$ = 8.7510Å, $3\sqrt{2}a_{STO}\cos(60^o)$).[13,29] This indicates that the $o$-GFO film is not able to show a cube-on-cube type growth on the unit-cell basis. However, if the $o$-GFO film growth is proceeded by the formation and deposition of a supercell with the dimension of $16a_{GFO} \times 18b_{GFO}$, the lattice mismatch between the GFO layer and the substrate surface can be effectively removed (with $\Delta_a = 0.01\%$, $\Delta_b = 0.08\%$). We will examine this proposition by looking into scanning transmission electron microscopy (STEM) images.

A bright-field STEM image is shown in Figure 2a for the cross-section of the [001]-oriented GFO film grown on a SRO(111) buffered STO(111) substrate. According to this cross-sectional mage, the thicknesses of the GFO layer and the SRO electrode are 200 nm and 30 nm, respectively. The selected area electron diffraction (SAED) pattern shown in Figure 2b is indexed by considering the superposition of the diffracted peaks along the zone axes $[010]$ and $[3\bar{1}0]$ of GFO and the peaks along the STO $[1\bar{1}0]$ zone axis (red rectangle) used as the standard. The diffracted peaks along the GFO $[010]$ zone axis (yellow rectangle) correspond to the $D_1$ domain configuration (Figure 1). On the other hand, the peaks along the GFO $[3\bar{1}0]$ zone axis (white rhombus) correspond to the $D_2$ domain configuration. The SAED patterns also indicate an epitaxial growth of the GFO film with the polar axis along the growth direction. One can notice two separated peaks by carefully examining the area encircled by green dots. This demonstrates that the GFO film is fully relaxed, rather than epitaxially strained.

High-resolution high-angle annular dark-field (HAADF) STEM images clearly show structural domain boundaries that are formed by distinct in-plane orientations. A HAADF STEM image of the area surrounded by a yellow rectangle is magnified and presented in the right-hand side of Figure 2c to clearly visualize the structural domain boundary. The three fast Fourier transformed images (upper regions of the magnified STEM image) indicate that the in-plane orientation of the central domain is different from those of the two neighboring domains. Figure 2c indicates that the width of the central domain is ~ 8 nm which nearly coincides with the *a*-axis dimension of the supercell proposed previously ($16a_{GFO} \times 18b_{GFO}$,). Thus, the central domain in Figure 2c represents the $D_1$ domain configuration, as depicted in Figure 1.

**PFM Images of the Polar c-axis-grown Film**

Vertical-mode piezoresponse force microscopy (vPFM) is a suitable method of measuring ferroelectricity or piezoelectricity for a small local region. In Figure 3a and b, we respectively present vPFM amplitude and phase images acquired over the area of 1$\mu$m×1$\mu$m. The phase contrast image shown in Figure 3b can be interpreted as the existence of ferroelectric domains with two antiparallel polarizations. However, this interpretation would be false if these domains were not remained at a particular polarization state (i.e., remain either up or down state) after turning off the bias *E*-field which had been used for the polarization switching. To clearly resolve this critical issue, we have chosen a particular region of the *o*-GFO film and applied an alternative *dc* voltage from +10 V to –10 V. Thus, the corresponding *E*-field is $\pm$500 kV/cm. The two applied voltages and the corresponding regions are marked in Figure 3c and d for the amplitude and phase contrast images, respectively. The two lower resolution vPFM images (10$\mu$m×10$\mu$m) shown in Figure 3c and d were taken immediately after turning off these *dc* voltages. Figure 3d clearly indicates that the 180º phase-shifted domains by the bias *E*-field return to the initial unshifted state as soon as the bias *E*-field is turned off. Thus, the present vPFM results do not show any evidence of ferroelectricity with a non-zero remanent polarization up to a bias *E*-field of $\pm$500 kV/cm.

**Ferroelectric Polarization Switching of the Polar *c*-axis-grown Films**

Having failed in obtaining a clear evidence of the ferroelectricity up to $\pm500$ kV/cm, we have then carried out *P-E* hysteresis measurement by applying much stronger *E*-fields, $E_{max}$ between 2500 and 4500 kV/cm. This has been done as our optimally processed [001]-oriented GFO/SRO/STO(111) film capacitors are characterized by the I-V current density as low as ~$10^{-6}$ A/cm$^2$ even at $\pm500$ kV/cm (i.e., at $\pm10$V; See Supplementary Information for details). This low leakage current is a remarkable improvement over those of other reported GFO films. For instance, the current density of the previously reported GFO/SRO/STO(111) film capacitor (with 0 % of MgO doping) is in the order of a few A/cm$^2$ for the same thickness of 200 nm (i.e., at $\pm500$ kV/cm).[32] Thus, our film capacitor processed by the optimized layer-by-layer growth of SRO shows a remarkable $10^6$-times improvement in the leakage-current density ($10^{-6}$ *vs.* $10^0$). Similarly, the reported current density of the GFO/Pt/YSZ(111) film capacitor at 300 K is as high as $1.4 \times 10^{-2}$ A/cm$^2$ (as estimated using 300 *μm* as the diameter of the top electrode) for the same thickness of 200 nm (i.e., at $\pm500$ kV/cm).[33] Again, our film capacitor shows a remarkable ~$10^4$-times improvement in the leakage-current density.

The Pt/*o*-GFO(001)/SRO(111) film capacitor grown on a STO(111) shows ferroelectric switching behavior with a remarkably high coercive field ($E_c$) of $\pm1400$ kV/cm at room temperature [Figure 4a]. Thus, $E_{max}$ value used in the PFM measurements is only 1/3 of the minimum electric-field needed for the polarization switching ($E_c$). Thus, the ferroelectric polarization switching cannot be attained by PFM techniques (with $E_{max}$ of $\pm10$ V) though a 180$^o$ phase-shift of the piezoresponse is readily observed. As shown in the inset of Figure 4a, the $P_r$ tends to reach its saturated value with increasing value of $E_{max}$. Thus, the net switching polarization ($2P_r$) estimated from Figure 4a is ~35 $\mu$C/cm$^2$ which is much bigger than the previously reported values of $2P_r$ (between 0.1 and 1.0 $\mu$C/cm$^2$).[14-16]

A more reliable value of the switching polarization can be obtained by employing PUND (positive-up & negative-down) pulse test. Figure 4b presents the PUND result of the 200-nm-thin GFO film capacitor obtained using the pulse delay time of 100 msec. We have found that the switching polarization decreases with increasing delay time and reaches a saturated value for the pulse delay time longer than ~80 msec. The net switching polarization ($2P_r$) is evaluated using the following relation: $2P_r = (\pm P^*) - (\pm P^\wedge)$. The $2P_r$ value obtained from Figure 4b is ~26 $\mu$C/cm$^2$. Comparing this value with the $2P_r$ value obtained from Figure 4a, one can conclude that the *P-E* curves overestimate the $2P_r$ value substantially. This suggests that the *ac* frequency of 1 kHz used in the *P-E* measurement is not high enough to completely eliminate the responses of mobile space charges which tend to be significant at lower *ac* frequencies.

To examine universality of the present finding of a large $2P_r$, we have also examined the polarization switching characteristics of the polar *c*-axis-oriented GFO film grown on a conducting ITO buffered cubic YSZ(001) substrate. As shown in Figure 5a, the GFO film (the same 200-nm thickness) grown on an ITO(001)/YSZ(001) substrate is characterized by the net switching polarization of ~30 $\mu$C/cm$^2$ at 298 K. As compared with the Pt/GFO/SRO/STO(111) capacitor (Figure 4a), the *P-E* curves of the Pt/GFO/ITO/YSZ(001) capacitor (Figure 5a) tend to have a more noticeable electrical-leakage problem. The I-V current density data supports this tendency of the electrical leakage in the Pt/GFO/ITO/YSZ(001) capacitor (See Figure S3b of Supplementary Information). The PUND result shown in Figure 5b also demonstrates that the Pt/GFO/ITO/YSZ(001) capacitor (with a 200-nm-thin GFO layer) is ferroelectric with the net switching polarization of ~20 $\mu C/cm^2$ at 298 K.

Compared with the $2P_r$ of the Pt/GFO/SRO/STO(111) capacitor, the $2P_r$ of the Pt/GFO/ITO/YSZ(001) capacitor is thus substantially reduced: ~35 $\mu C/cm^2$ *vs.* ~30 $\mu C/cm^2$ according to their *P-E* results. As stated previously, the *c*-axis-grown *o*-GFO film on an ITO/YSZ(001) substrate is characterized by the six crystallographic variants having total twelve in-plane orientations with a successive in-plane tilting angle of 30°, in spite of little lattice mismatch between the GFO film and the YSZ substrate (See Supplementary Information for details). The reduced $2P_r$ value can possibly be correlated with this increase in the crystallographic variants (*i.e.*, the degree of complexity in the domain configuration) upon replacing the SRO/STO(111) substrate with the ITO/YSZ(001) substrate. However, further in-depth studies should be made before clearly resolving the origin of the substrate-dependent $2P_r$.

**DISCUSSION**

For in-depth understanding of the atomic-scale origin of the observed polarization switching, we have calculated the DFT polarization of the polar $Pna2_1$ phase of *o*-GFO using the Berry-phase method.[34,35] In order to obtain a correct evaluation of the ferroelectric polarization, a centrosymmetric prototypic phase of GFO should be first identified. For doing this, we have utilized the PSEUDO code of the Bilbao crystallographic server,[36] which lattice dynamically allows one to determine the nearest supergroup structure for an input arbitrary structure. In the present case, four centrosymmetric supergroup structures are examined. They are *Pnna, Pccn, Pbcn*, and *Pnma*. Among these four, the nearest reference structure, in terms of the total displacement of all atoms, is the *Pnna* phase, which is consistent with the previously reported result.[20] The optimized lattice parameters of the ferroelectric $Pna2_1$ and prototypic *Pnna* phases were subsequently obtained by calculating the Kohn-Sham (K-S) energy as a function of the unit-cell volume and finding its minimum which corresponds to the ground-state K-S energy in the absence of any external pressure. The optimized lattice parameters are: (i) $a = 5.1647$ Å, $b = 8.8197$ Å, and $c = 9.5079$ Å for the *Pna*2$_1$ structure, and (ii) $a = 5.2672$ Å, $b = 8.9193$ Å, and $c = 9.5684$ Å for the *Pnna* structure. The optimized structures of the *Pnna* and *Pna*2$_1$ phases of o-GFO are depicted in Figure 6a.

According to group theoretical analysis,[37] there exists only one conceivable transition path that connects the prototypic *Pnna* phase to the ferroelectric *Pna2$_1$* phase. We have decomposed the atomic displacements that relate the nonpolar *Pnna* phase to the polar *Pna2$_1$* phase into the symmetry-adapted mode of the prototypic phase. The resulting symmetry-adapted mode is exclusively given by $\Gamma_4^-$. Thus, the *Pnna*–to–*Pna2$_1$* phase transition should be mediated by the freezing-in of the zone-center $\Gamma_4^-$ polar phonon. Let us define the displacement amplitude of the polar $\Gamma_4^-$ phonon as $Q_{\Gamma_4^-}$. In Figure 6b, the K-S energy and the Berry-phase polarization are plotted as a function of the mode amplitude $Q_{\Gamma_4^-}$.[38] Here, the polarization [lower panel of Figure 6b] is given by the product of $Q_{\Gamma_4^-}$ and the Born effective charge tensor. As shown in Figure 6b, the computed K-S energy exhibits a double-well-type potential, which demonstrates the relative stability of the ferroelectric *Pna*2$_1$ phase over the prototypic nonpolar *Pnna* phase with the energy difference of 1.05 eV per formula unit (f.u.). Herein the equilibrium ferroelectric polarization of the *Pna*2$_1$ phase is given by the computed polarization values (lower panel) at the two K-S energy minima, namely, $\pm 25.67$ μC/cm$^2$, which corresponds to $Q_{\Gamma_4^-}$ of $\pm 1$, respectively. Since the polarization switching is expected to occur along the *Pna2$_1$*–to–*Pnna* phase-transition path, i.e., $\Gamma_4^-$, the activation free-energy of the polarization switching between the double wells can be obtained from Figure 6b, which is ~1.05 eV per f.u. This value lies within the two extreme *ab initio* values (0.52 and 1.30 eV per f.u.) previously obtained by Stoeffler[20] using ABINIT & FLAPW/ FLEUR codes. The activation barrier of 1.05 eV per f.u. is about 2.5 times bigger than that of BiFeO$_3$, the most extensively studied multiferroic, and 20 times bigger than that of Pb(Zr,Ti)O$_3$, the most widely used displacive ferroelectrics.[39] This unusually high activation barrier indicates that the polar *Pna*2$_1$ phase in *o*-GFO is very stable against thermally activated random dipole switching across the centrosymmetric *Pnna* barrier. The observed high $E_c$ ($\pm$1400 kV/cm; Figure 4a) can also possibly be correlated with this high activation barrier.

Finally, we will correlate the reported high ferroelectric transition temperature (> 1368 K; ref. 22) with this remarkably high activation free-energy. According to the transition-state theory of rate processes,[40] the frequency ($\nu$) of the polarization switching across the *Pnna* potential barrier can be written as

$$\nu = \frac{k_B T}{h} e^{-\Phi_o/k_B T} \qquad (1)$$

where $\Phi_o$ is the barrier height of the polarization switching which can be treated as the difference in the Kohn-Sham energy between *Pnna* and *Pna*2$_1$ phases (1.05 eV per f.u.). Strictly speaking, Equation (1) is valid for describing the dipole-switching rate of order-disorder ferroelectrics. For sufficiently high temperatures near the phase-transition point ($T_c$), however, Equation (1) is also applicable to the dipole-switching rate of displacive ferroelectrics[41] that include the present *Pna2$_1$*–to–*Pnna* transition. As temperature increases and approaches $T_c$, the dipole-switching rate becomes so rapid that the mean residence time ($\tau_o$) of the bound polarization in one of the two ferroelectric double wells becomes shorter than a certain critical time for the experimental observation. Under this condition, the net ferroelectric polarization effectively disappears because of the switching average of two opposite polarizations across the centrosymmetric barrier [*Pnna* state in the present case; Figure 6b]. Let $\nu_o$ be the frequency at which the net bound polarization just begins to disappear. In the vicinity of $T_c$, $\Phi_o < k_B T$ for a fixed value of $\Phi_o$. Under this condition, $\nu_o$ can be approximated by the following expression:

$$\nu_o = \frac{1}{\tau_o} \approx \frac{k_B T_o}{h}\left(1 - \frac{\Phi_o}{k_B T_o}\right) \quad (2)$$

where $T_o$ denotes the temperature that corresponds to the critical switching frequency $\nu_o$. Since $T_c$ is expressed by $T_c = T_o + \epsilon$, where $\epsilon$ is a small positive number (including zero), $T_c$ can be correlated with $\Phi_o$ by the following relation for a fixed value of $\Phi_o$:

$$T_c = \frac{(h\nu_o + \Phi_o)}{k_B} + \epsilon \approx \frac{(h\nu_o + \Phi_o)}{k_B} \quad (3)$$

Equation (3) clearly indicates that $T_c$ is linearly proportional to $\Phi_o$. This explains the observed high $T_c$ of *o*-GFO in terms of the unusually high activation free-energy barrier ($\Phi_o$) as predicted by *ab initio* calculations.

**CONCLUSIONS**

We have clarified a puzzling discrepancy between the observed $P_r$ values ($\leq 0.5$ $\mu C/cm^2$) and the *ab initio* prediction and have unequivocally demonstrated the ferroelectric polarization switching with the net switching polarization of ~30 $\mu C/cm^2$ by using the two distinct types of *o*-GFO thin films preferentially grown along the polar *c*-axis in *Pna*2$_1$ setting (*b*-axis in *Pc*2$_1$*n* setting). The estimated activation energy for the polarization switching is ~1.05 eV per f.u. This high value accounts for the observed stability of the polar *Pna*2$_1$ phase over the centrosymmetric *Pnna* phase for a wide range of temperature up to 1368 K.

**CONFLICT OF INTEREST**

The authors declare no conflict of interest.

**ACKNOWLEDGEMENTS**

This work was supported by the National Research Foundation (NRF) Grants funded by the Korea Government (MSIP) (Grant No. 2012R 1A1A2041628 & 2013R 1A2A2A01068274). The work at Cambridge was supported by the Engineering and Physical Sciences Research Council (EPSRC).

**\* Figure Captions**

**Figure 1** X-ray diffraction (XRD) data and in-plane domain orientations. **(a)** $\theta$-$2\theta$ XRD pattern of the [001]-oriented 200-nm-thin GFO thin film grown on a SRO(111) buffered hexagonal STO(111) substrate. **(b)** Schematic representation of the most probable domain configuration for the polar *c*-axis-oriented GFO film grown on a STO (111) substrate. Blue bi-directional arrows refer to in-plane orientation of the STO(111) substrate. **(c)** In-plane XRD $\varphi$-scan spectra of the polar *c*-axis grown GFO film, as obtained by keeping the Bragg angle at (013) for *o*-GFO film (orange line) and at (110) and (100) for STO (blue and green lines, respectively). A schematic diagram presented below the $\varphi$-scan spectra denotes the projected normal vectors of (013) for GFO on STO(111) and projected normal vectors of (110) and (100) STO on STO(111). **(d)** A schematic domain configuration of the polar *c*-axis-oriented GFO film grown on a STO(111) substrate showing six in-plane domain orientations with three distinct crystallographic variants. Three different kinds of the projected normal vectors are shown in the lower corner.

**Figure 2** STEM images of the polar *c*-axis-oriented GFO film. **(a)** A cross-sectional BF-STEM image of the polar *c*-axis-oriented GFO film grown on a SRO(111) buffered hexagonal STO(111) substrate. **(b)** A SAED pattern confirming the in-plane orientation between the GFO film and the STO substrate. The diffraction pattern is indexed by considering the superposition of the diffracted peaks along the zone axes [$3\bar{1}0$] and [010] of GFO and the zone axis [$1\bar{1}0$] of STO. **(c)** A HAADF STEM image for the interfacial region between the polar *c*-axis-oriented GFO film and the SRO electrode layer. The area

surrounded by a yellow rectangle is magnified in the right-hand side to clearly visualize the structural domain boundary. Three fast Fourier transformed images are also shown in the upper region of the magnified STEM image.

**Figure 3** Vertical-mode PFM images of the polar *c*-axis-oriented GFO film. The GFO film was grown on a SRO(111) buffered STO (111) substrate. PFM **(a)** amplitude and **(b)** phase contrast images of the GFO film acquired over the area of 1$\mu$m×1$\mu$m. PFM **(c)** amplitude and **(d)** phase contrast images of the two concentric square regions obtained after applying an alternative *dc* voltage from +10 V to –10 V and subsequently turning off these *dc* voltages.

**Figure 4** Ferroelectric responses of the GFO film grown on a hexagonal STO substrate. **(a)** *P-E* curves of the Pt/GFO(001)/SRO(111) film capacitor grown on a STO(111) showing ferroelectric switching behavior with a high coercive field ($E_c$) of ±1400 kV/cm at 298 K. **(b)** PUND test result of the polar *c*-axis-oriented 200-nm-thin GFO film capacitor grown on a SRO(111)/STO(111) obtained using the pulse delay time of 100 msec.

**Figure 5** Ferroelectric responses of the GFO film grown on a cubic YSZ substrate. **(a)** *P-E* curves of the Pt/GFO(001)/ITO(001) film capacitor grown on a YSZ(001). The coercive field ($E_c$) deduced from these curves is ±1100 kV/cm at 298 K. **(b)** PUND test result of the polar *c*-axis-oriented 200-nm-thin GFO film capacitor grown on an ITO(001)/YSZ(001).

**Figure 6** Crystal structures and the associated ferroelectric double-well potential. **(a)** Optimized unit-cell crystal structures of prototypic *Pnna* and polar *Pna*2$_1$ phases of *o*-GFO

obtained after structurally relaxing the computed Kohn-Sham energy as a function of the unit-cell volume. Eigenvectors of the $\Gamma_4^-$ phonons at Ga and Fe ions are shown in the right-hand side using blue arrows. **(b)** The computed Kohn-Sham energy plotted as a function of the fractional amplitude $Q_{\Gamma_4^-}$. The reference state of $Q_{\Gamma_4^-} = 0$ denotes the prototypic *Pnna* phase. In the lower panel, the computed polarization values are plotted as a function of $Q_{\Gamma_4^-}$.

# * Figures

# Figure 1

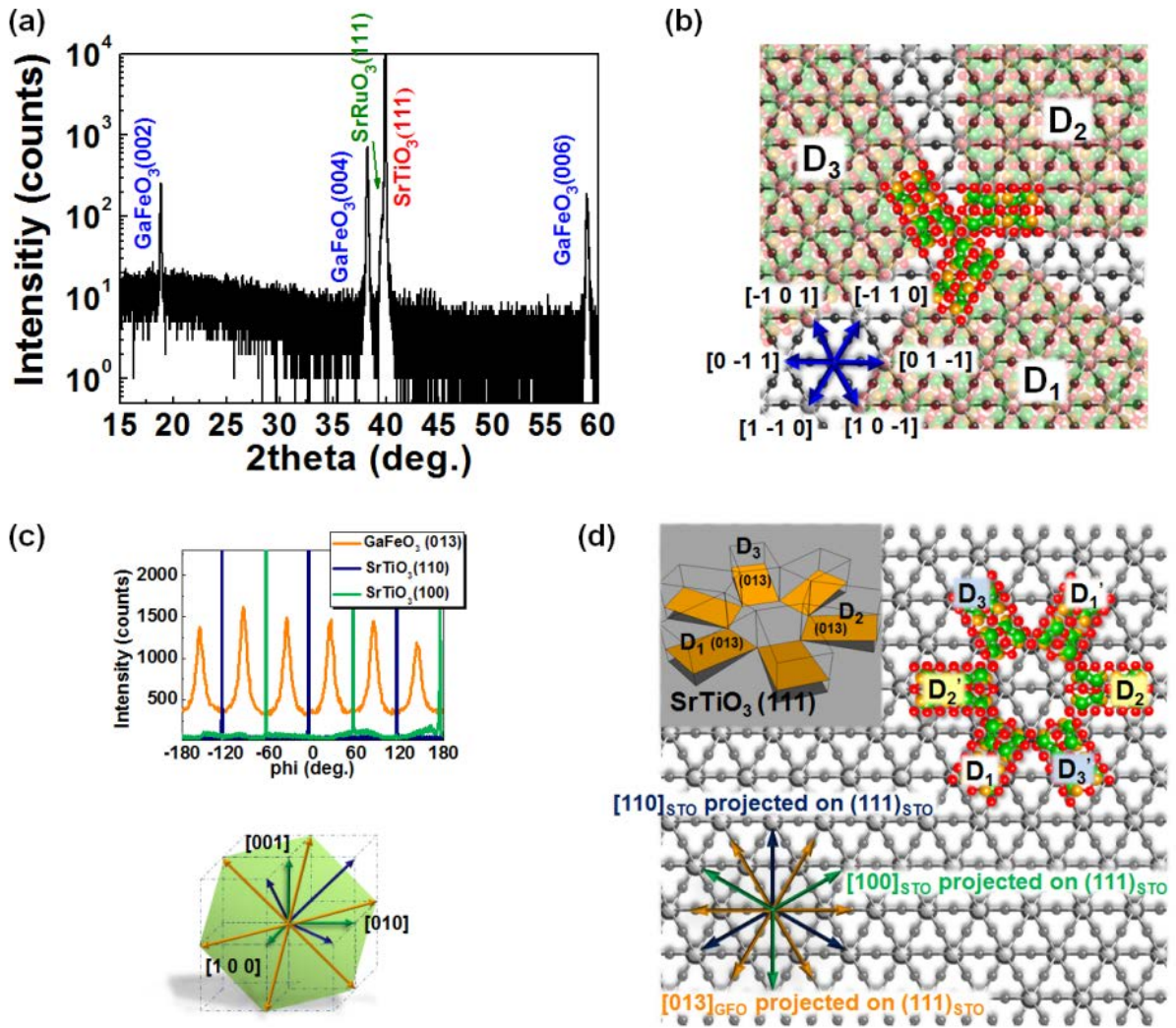

**Figure 2**

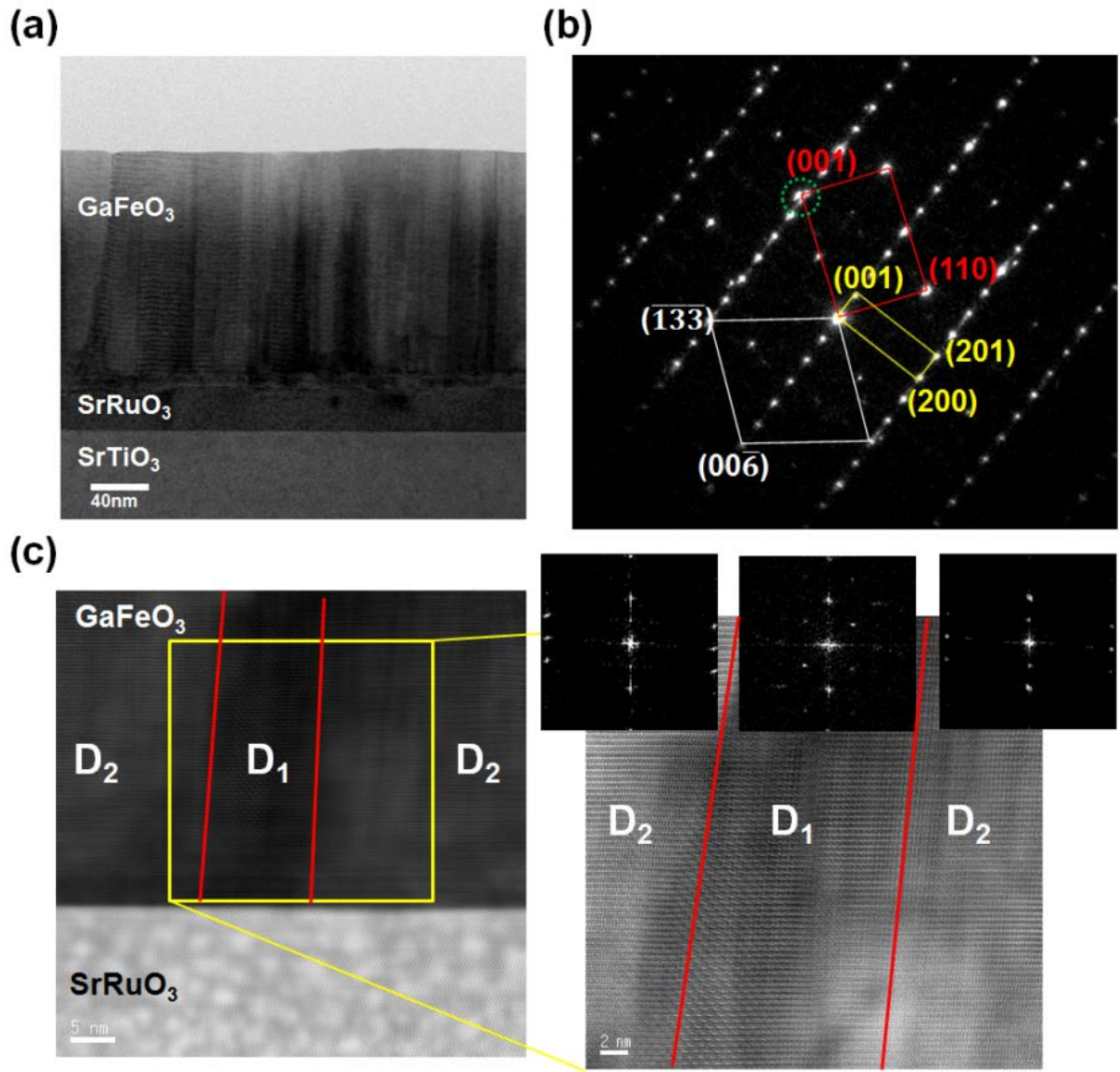

**Figure 3**

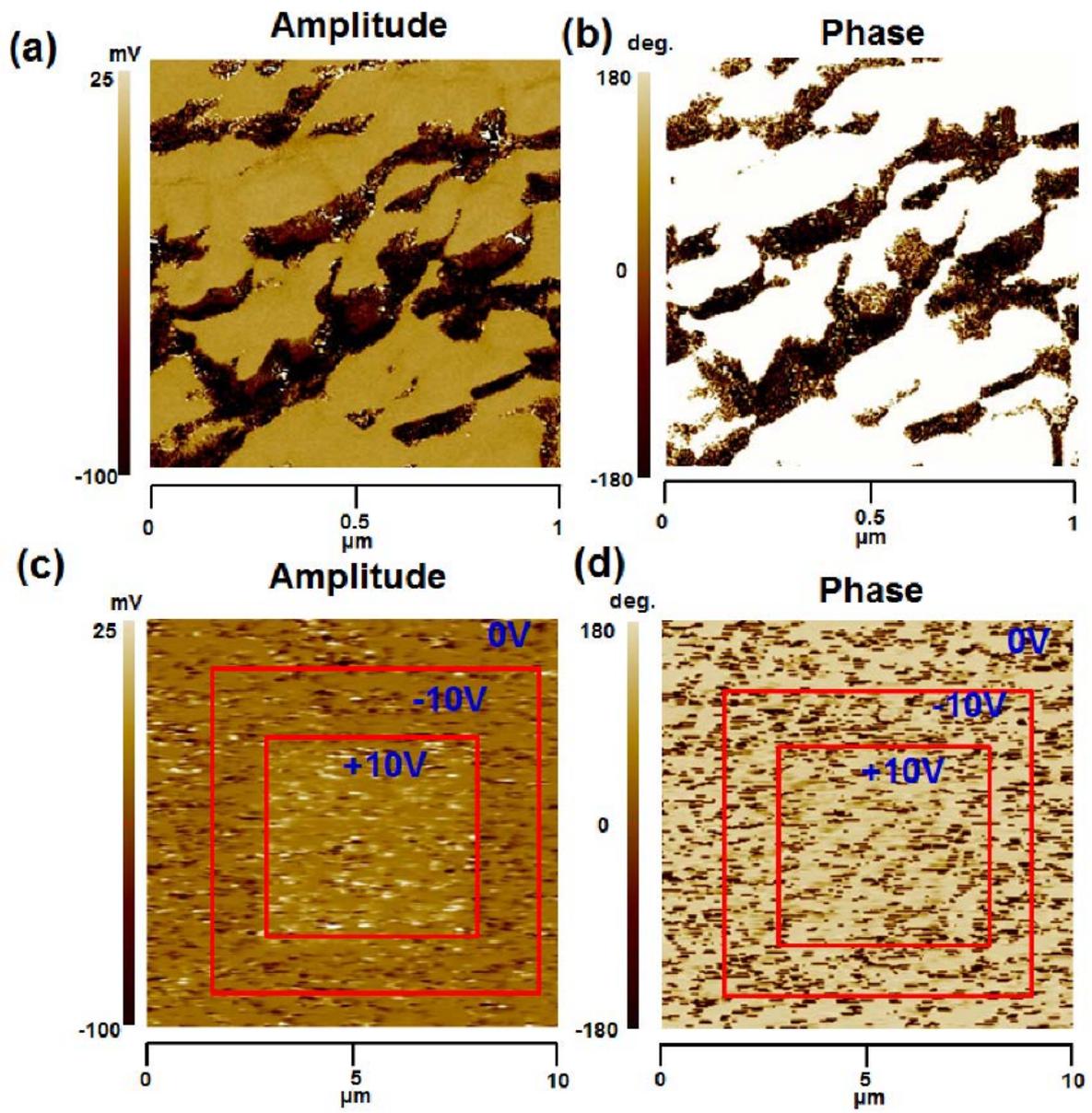

**Figure 4**

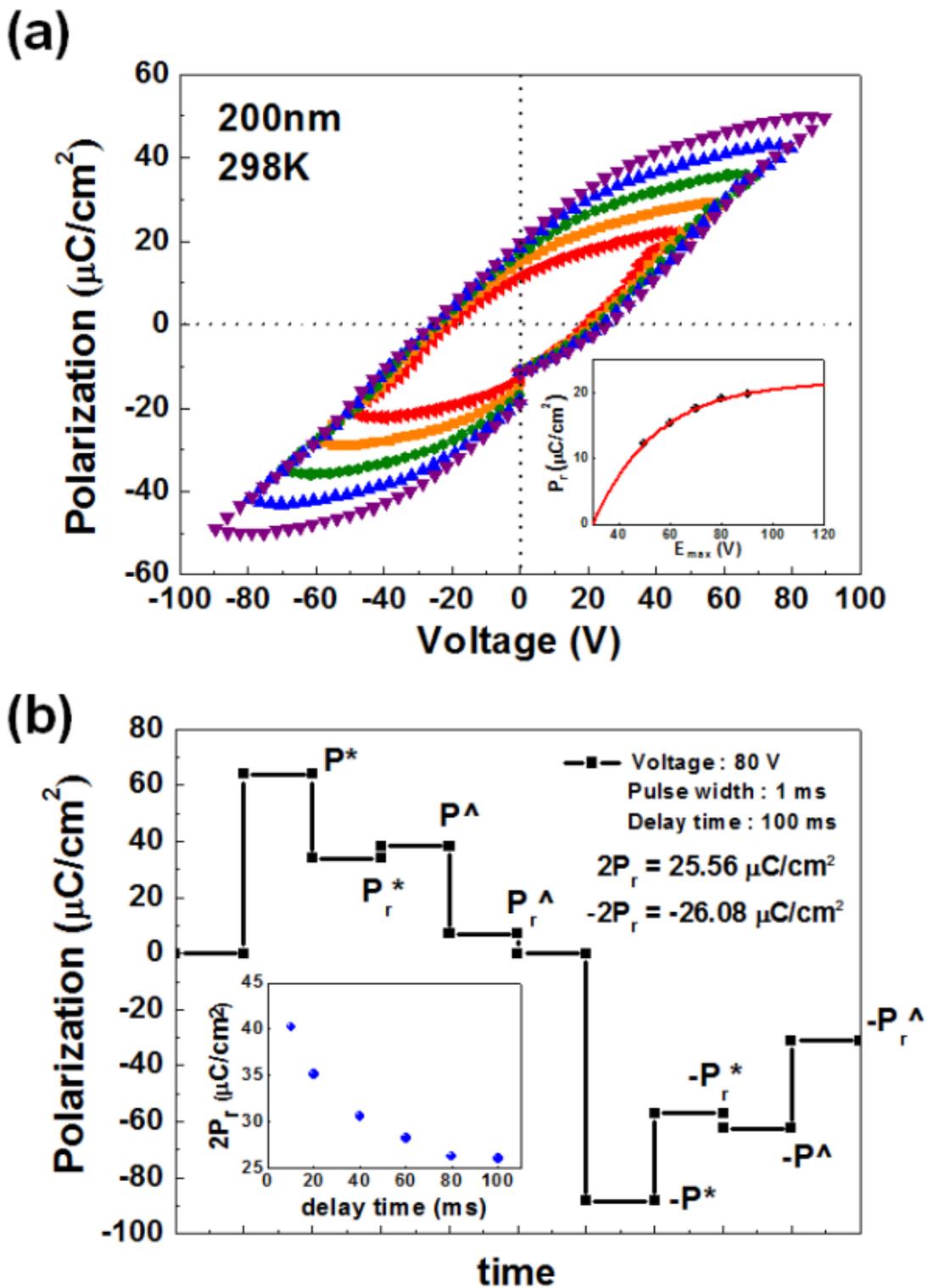

**Figure 5**

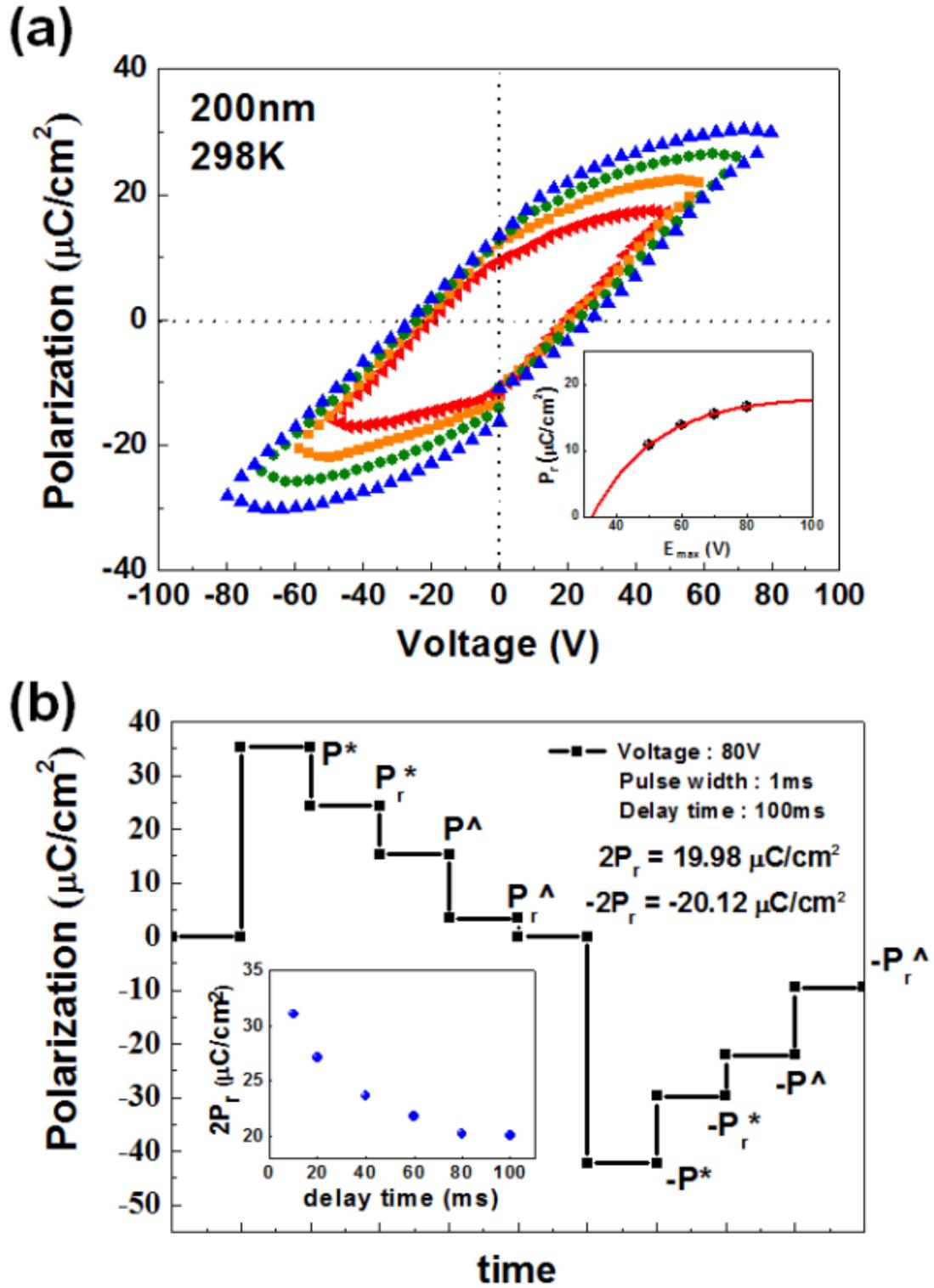

**Figure 6**

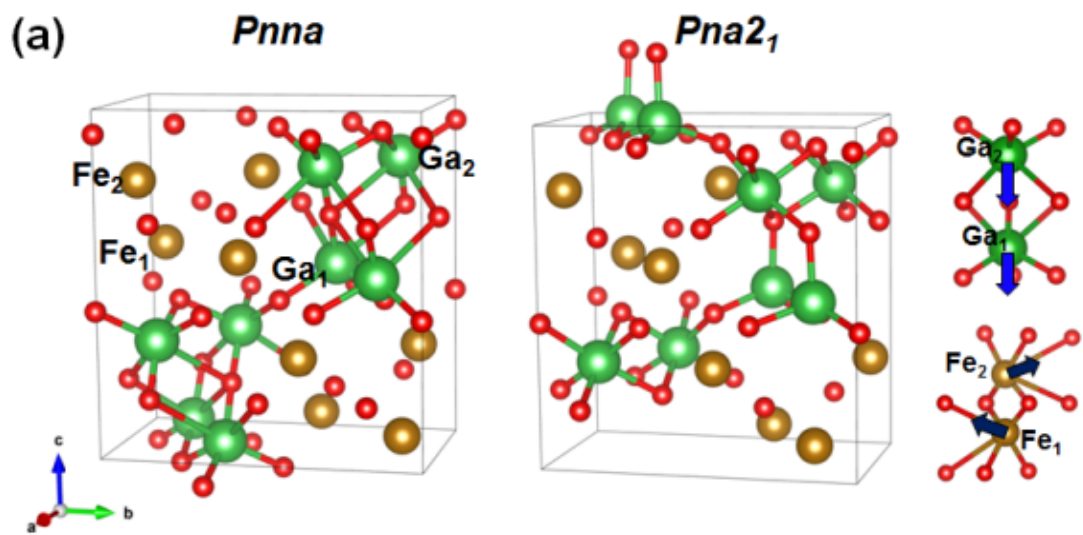

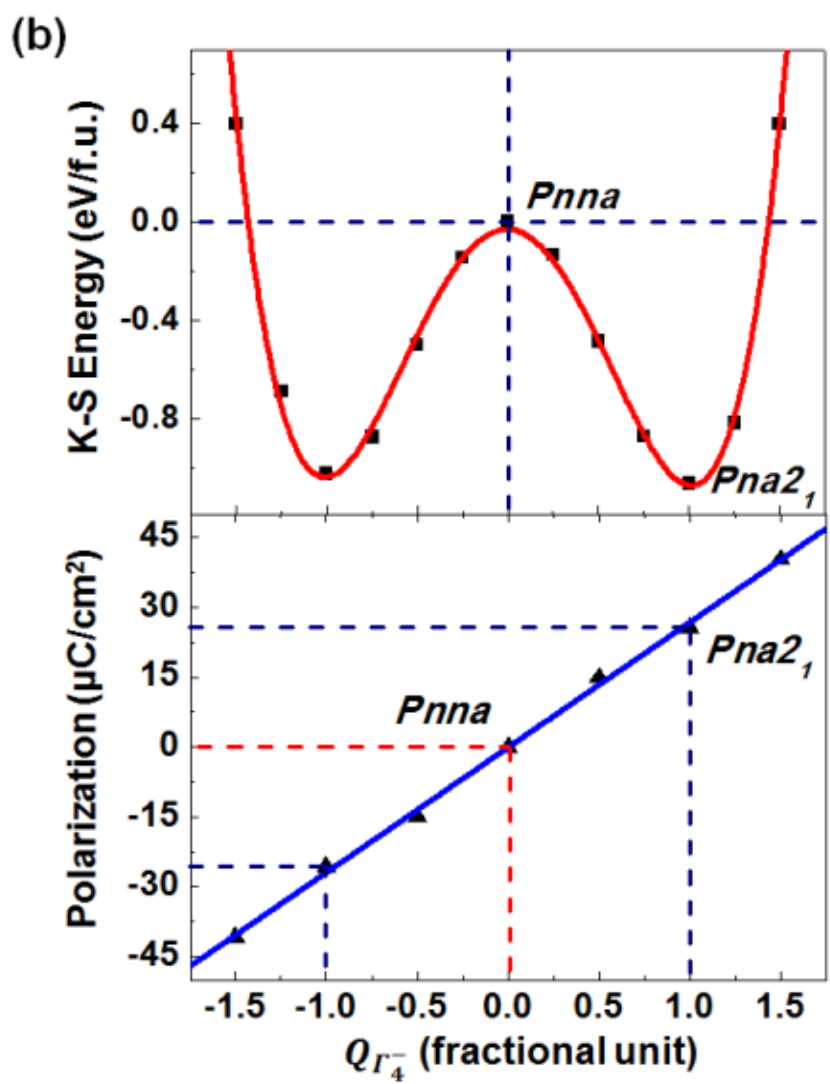

SUPPLEMENTARY INFORMATION

# Ferroelectric polarization switching with a remarkably high activation-energy in orthorhombic GaFeO$_3$ thin films

Seungwoo Song,[1] Hyun Myung Jang,[1*] Nam-Suk Lee,[2] Jong Y. Son,[3] Rajeev Gupta,[4] Ashish Garg,[4] Jirawit Ratanapreechachai,[5] and James F. Scott[5,6+]

[1]Division of Advanced Materials Science (AMS) and Department of Materials Science and Engineering, Pohang University of Science and Technology (POSTECH), Pohang 790-784, Republic of Korea.

[2]National Institute for Nanomaterials Technology (NINT), Pohang University of Science and Technology (POSTECH), Pohang 790-784, Republic of Korea.

[3]Department of Applied Physics, College of Applied Science, Kyung Hee University, Giheung-Gu, Yongin-City 446-701, Republic of Korea.

[4]Department of Materials Science and Engineering, Indian Institute of Technology (IIT), Kanpur, Kanpur 208016, India.

[5]Cavendish Laboratory, Department of Physics, University of Cambridge, Cambridge CB3 0HE, united Kingdom.

[6]School of Chemistry and School of Physics, University of St. Andrews, St. Andrews KY16 9ST, United Kingdom.

*Correspondence and requests for materials should be addressed to either H. M. J. (email: hmjang@postech.ac.kr ) or J. F. S. (email: jfs32@cam.ac.uk )

## 1. Magnetic properties

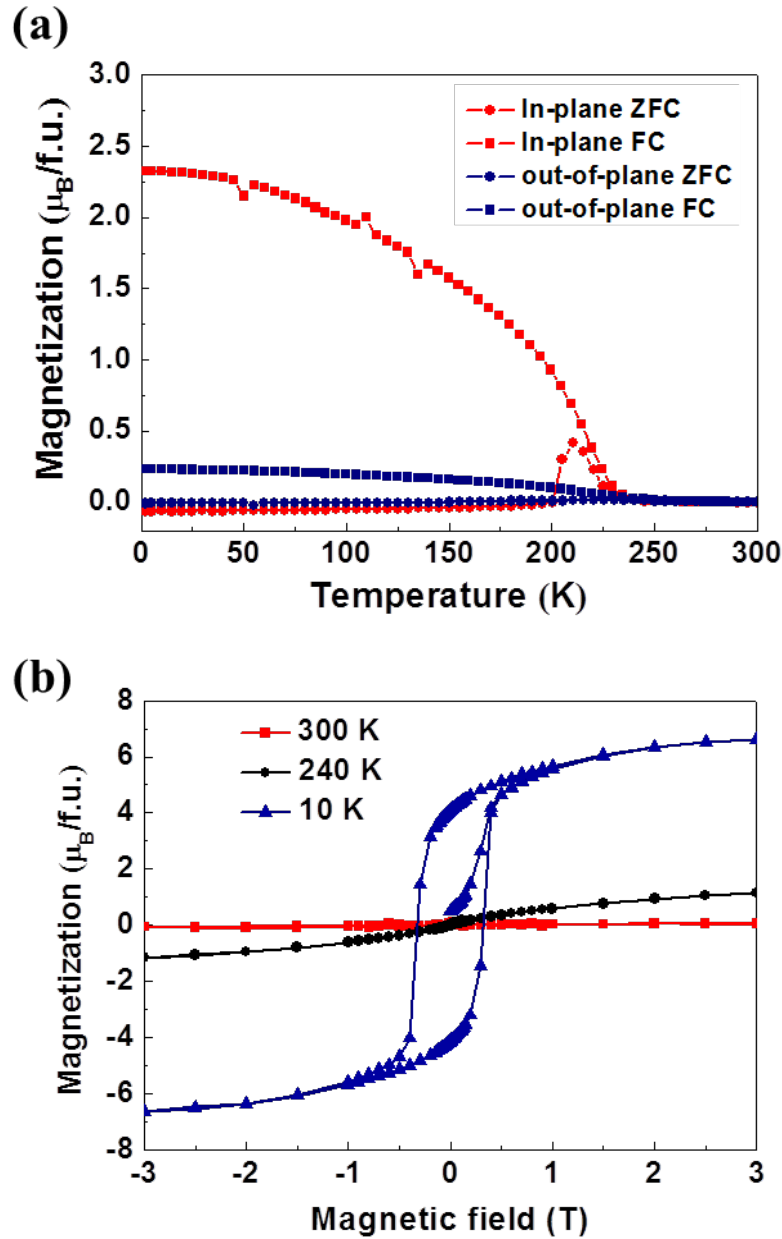

**Figure S1. (a)** Temperature-dependent field cooled (FC) and zero-field cooled (ZFC) magnetization curves of the polar *c*-axis-oriented *o*-GFO film (200-nm-thin) grown on a SRO buffered STO(111) substrate. Notice that a large discrepancy between the in-plane FC and ZFC curves for temperatures below ~230 K. This indicates that the magnetic easy axis lies along the in-plane direction (i.e., perpendicular to *c*-axis) and a (ferrimagnetic) ordering occurs at ~230 K. **(b)** Magnetization *versus* magnetic-field (M-H) hysteresis curves at three selected temperatures, indicating that the magnetic-ordering temperature is lower than 240 K.

## 2. In-plane domain orientation of GFO film grown on ITO/YSZ (001)

As presented in Figure S2(a), the x-ray $\phi$-scan spectra of both GFO {122} and {013} planes are characterized by 12 distinct peaks, and each peak is separated from the two neighboring peaks by 30°. Here, we have elucidated the detailed in-plane orientations of the GFO domains by performing x-ray $\phi$-scan experiments for both GFO {013} and {122} planes. From the three $\phi$-scan spectra shown in Figure S2(a), one can establish the following orientation relationships among GFO, ITO, and YSZ: The projection of the normal vector of YSZ (111) on YSZ (001) is nearly parallel to the projection of the normal vector corresponding to the peak A of GFO (122). On the contrary, the projection of the normal vector of YSZ (111) on YSZ (001) is rotated by 45° from that of GFO (013). These relationships can be readily identified by examining our schematic drawing of GFO (122) and (013) and YSZ (111) on YSZ (001) [Figure S2(b)]. The three projected normal vectors are summarized in the inset of Figure S2(b).

The domain orientation corresponding to the peak A of the *c*-axis grown GFO film [on YSZ (001)] is denoted by $D_1$ in Figure S2(b). On the other hand, the two additional domain orientations corresponding to the peaks B and C are denoted by $D_2$ and $D_3$, respectively. In Figure S2(b), $D_x'$ denotes the GFO domain which is rotated from the $D_x$-domain by 90 degrees. Thus, the observed 12 peaks can be unequivocally explained by the three distinct domains, $D_1$, $D_2$, and $D_3$, with a successive tilting angle of 30°, showing total twelve in-plane orientations.

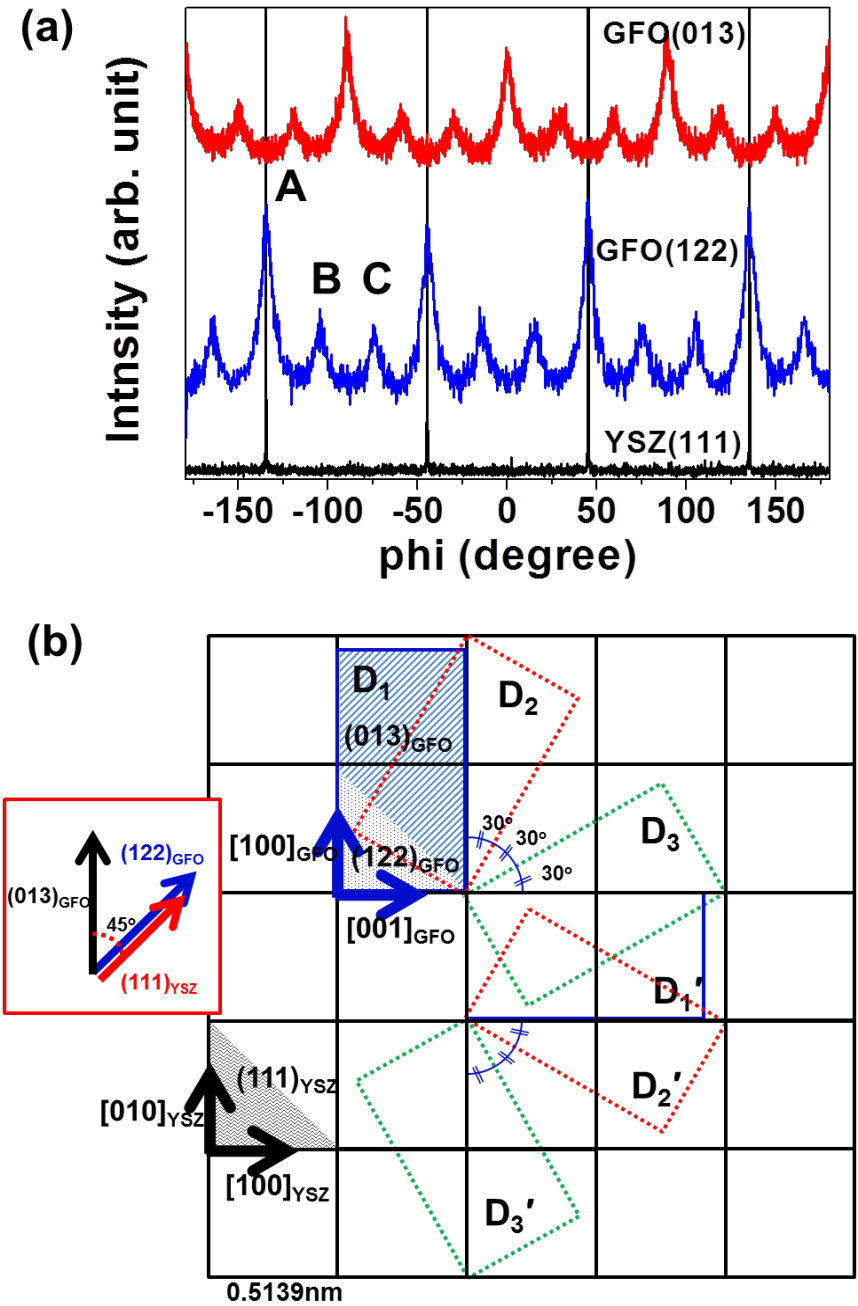

**Figure S2.** (a) X-ray $\phi$-scan spectra of {111} planes of YSZ (black line), {122} planes of GFO (blue line), and {013} planes of GFO (red line). (b) Schematic drawing of the domain orientations of the GFO film on YSZ (001). Herein, (111) plane of YSZ is shaded as grey, while (122) and (013) planes of the GFO film on YSZ (001) are shaded as pale blue and darker blue, respectively.

## 3. Current *versus* voltage (I-V) curve

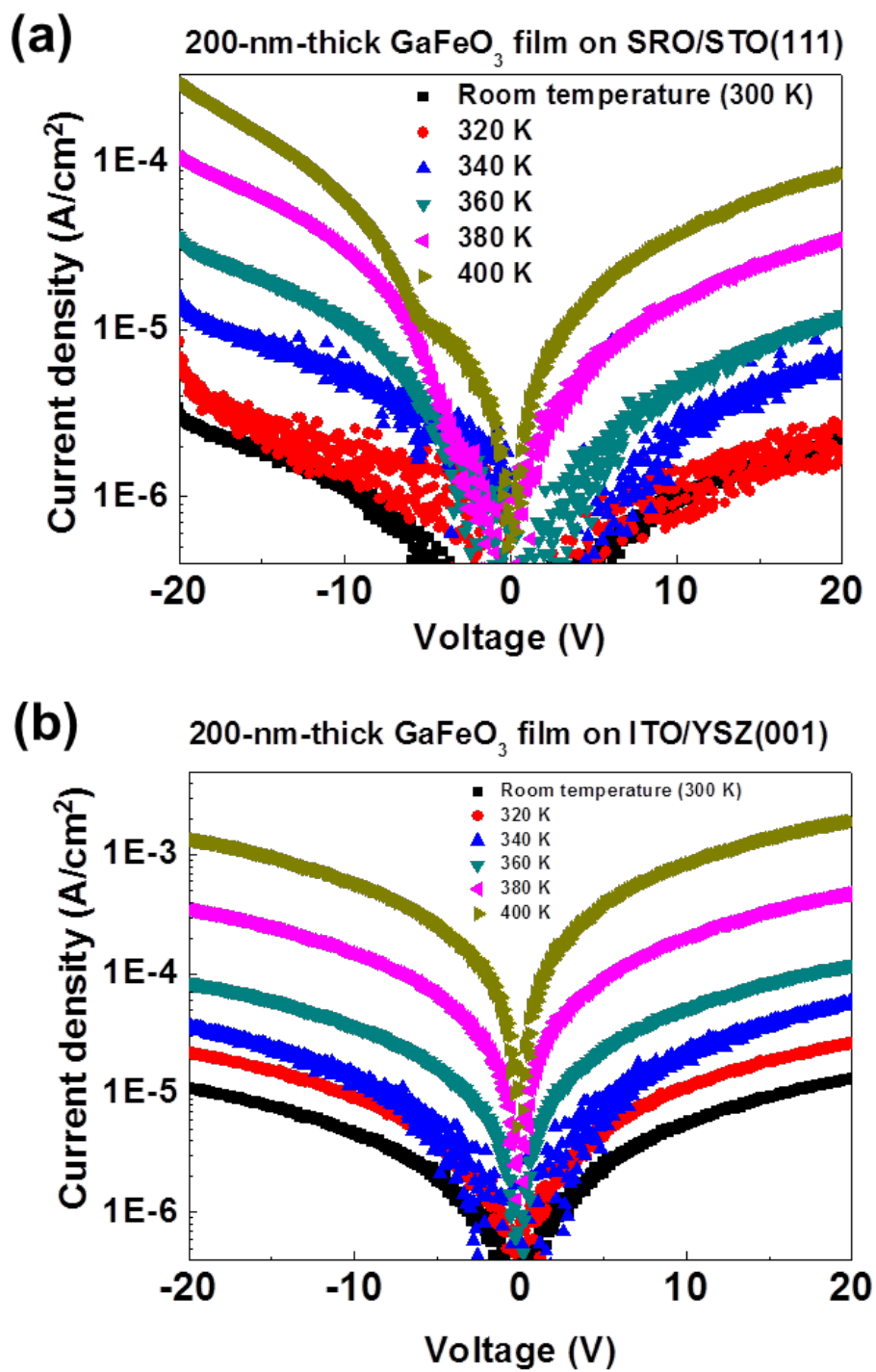

**Figure S3.** Temperature-dependent I-V curves of the polar *c*-axis-oriented *o*-GFO film (200-nm-thin) grown on **(a)** a SRO(111) buffered STO(111) substrate and **(b)** an ITO(001) buffered YSZ(001) substrate.

## 4. Atomic force microscopy (AFM) image of the substrate.

One side-polished STO(111) substrate was dipped into a hot, ultrasonically agitated buffered hydrogen fluoride (BHF) solution (NH$_4$F:HF = 7:1) for 30 seconds at 63°C. Then, the substrate was annealed in a tube furnace at 1050°C under oxygen ambient for 1 hour.

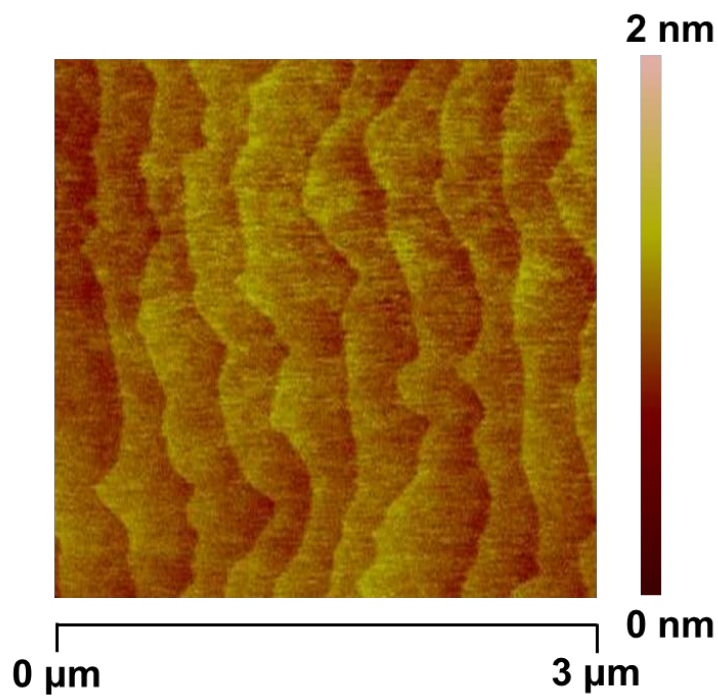

**Figure S4.** A topographic AFM image of the chemically and thermally treated STO(111) substrate.

## 5. AFM and XRD data of the SRO bottom electrode

Figure S5(a) shows XRD $\theta$-$2\theta$ scan of SRO/STO(111). One can notice the thickness fringes of the SRO(222) around the $2\theta$ of the STO(222), which is a strong evidence of the epitaxial growth. The positions of interference fringes displayed in the inset of Figure S5(a) were used to determine the film thickness which is estimated be 30 *nm*.

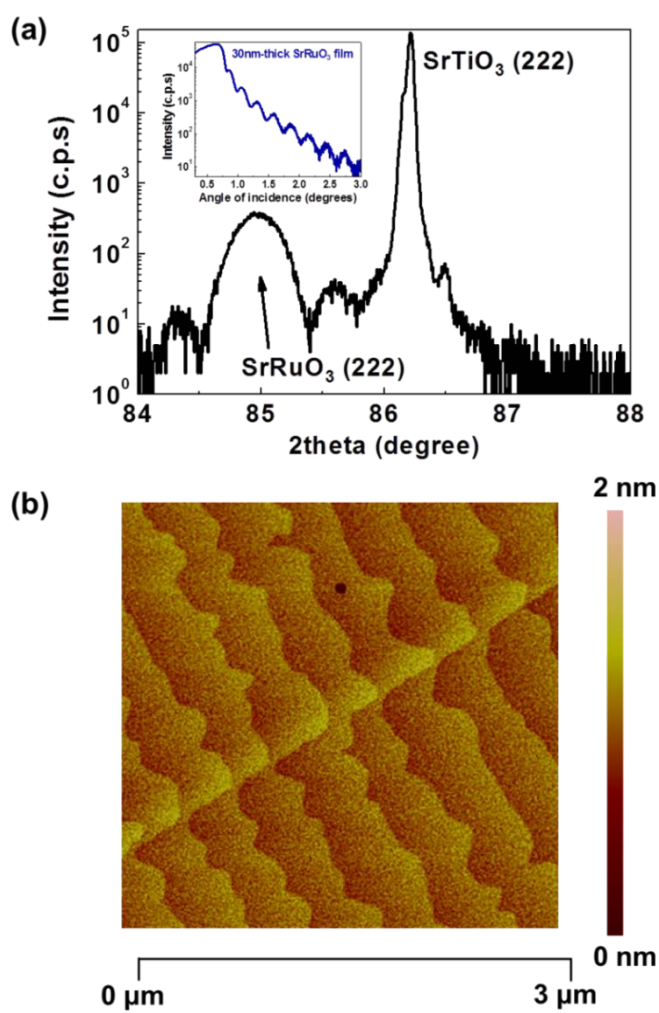

**Figure S5.** **(a)** XRD $\theta$-$2\theta$ scan of the SRO electrode layer on STO(111) with the fringe around $2\theta$ of the STO(222) peak. The inset shows x-ray reflectivity (XRR) data for the 30-nm-thick SRO bottom electrode layer. **(b)** An atomically flat AFM image of the SRO layer grown on a STO(111) substrate.

## 6. AFM image of the 200-nm-thick polar *c*-axis grown GFO film

Figure S6 displays topographic images of the polar *c*-axis-oriented GFO (200-nm-thick) film grown on a SRO(30nm)/STO(111) substrate during the PFM measurement.

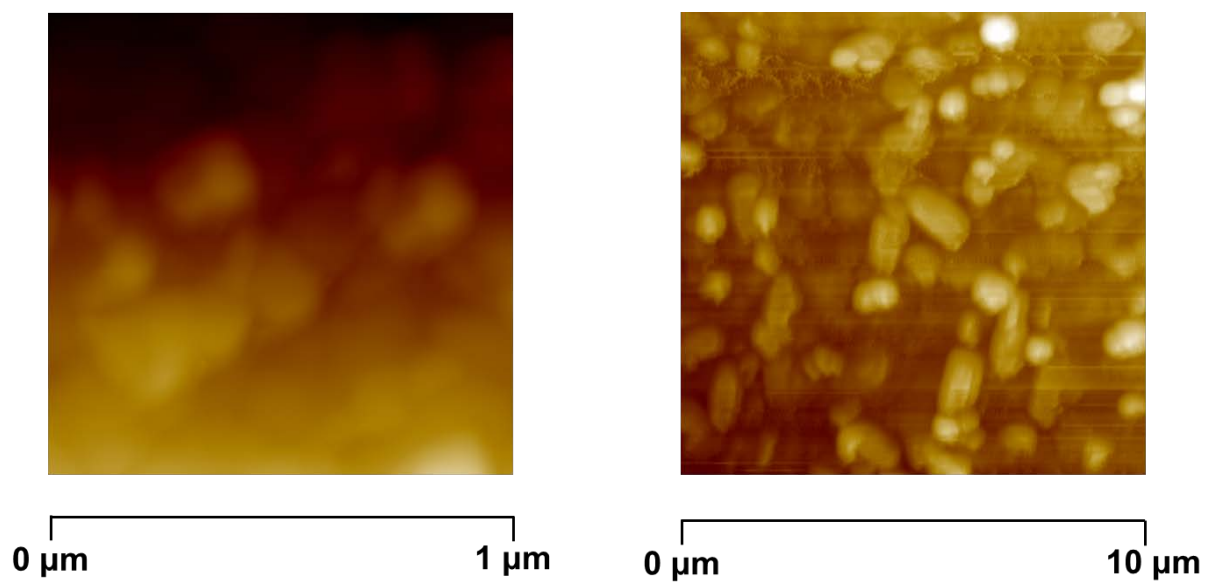

**Figure S6.** Atomically flat AFM images of the *o*-GFO film grown on a SRO(111) buffered STO(111) substrate.

# 7. Computational procedures for lattice dynamics study of the *Pnna-to-Pna2₁* structural phase transition

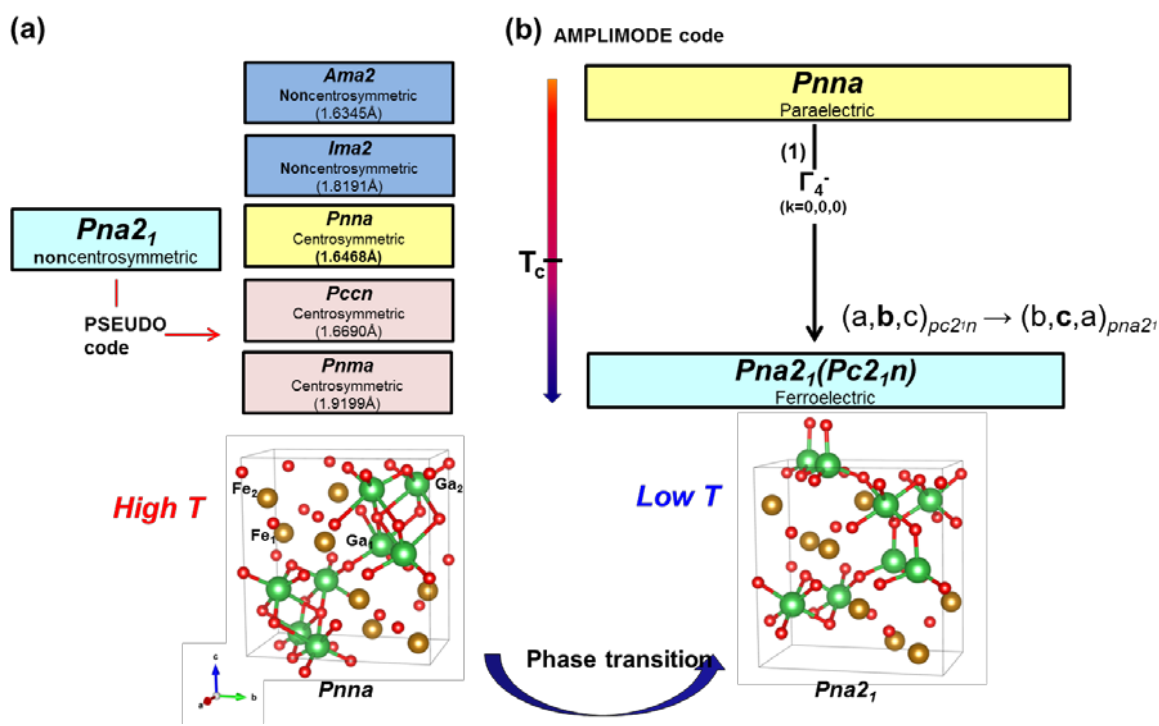

**Figure S7. (a)** Looking for a pseudo-symmetry among the supergroup structures for the input polar *Pna2₁* structure using the PSEUDO code of the Bilbao crystallographic server. Amomg these supergroup structures, *Pnna* is the most relevant paraelectric centrosymmetric structure that has the least atomic-displacement for each constituting atom. **(b)** AMPLIMODES calculations were carried out for a symmetry-adapted mode analysis of a displacive phase transition. Starting from the experimental structures of the high- and low-symmetry phases, the program determines the global structural distortion that relates to the two phases. The symmetry modes compatible with the symmetry break are then calculated. Their orthogonality permits the decomposition of the global distortion, obtaining the amplitudes of the different symmetry-adapted distortions present in the structure, as well as their corresponding polarization vectors. As indicated, the *Pnna-to-Pna2₁* structural phase transition in *o*-GFO is mediated solely by the freezing-in of the zone-center $\Gamma_4^-$ polar phonon.